\documentstyle[preprint,pra,aps,epsf]{revtex}

\begin{document}
\draft
\title{\bf Enhancement factor for the electron electric
dipole moment in francium and gold atoms}
\author{T. M. R. Byrnes, V. A. Dzuba, V. V. Flambaum,
and D. W. Murray}
\address{School of Physics, University of New South Wales,
Sydney, 2052, Australia}
\maketitle
\begin{abstract}
If electrons had an electric dipole moment (EDM) they would
induce EDMs of atoms. The ratio of the 
atomic EDM to the electron
EDM for a particular atom is called the enhancement factor, $R$.
We calculate the enhancement
factor for the francium and gold atoms, with
the results
$910 \, \pm \sim 5\%$ for Fr and $260 \, \pm \sim 15\%$ for Au.
The
large values of these enhancement factors make these atoms
attractive for electron EDM measurements, and hence the search
for time-reversal invariance violation. 
\end{abstract}
\vspace{5mm}
\vspace{10mm}

The existence of a non-zero electric dipole moment (EDM)
of an atom, electron, or any quantum-mechanical system would
imply that time-reversal invariance ($T$) and parity ($P$)
are violated. To date, no non-zero EDM has been observed,
though experimental limits on their magnitude have been
obtained.
The Standard Model predicts an upper limit on the
electron EDM of the order of
$10^{-40} e \mbox{ cm}$ \cite{PK91},
while various alternative
models predict values many orders of magnitude larger (for
reviews of predicted values of the electron EDM see
Refs.\ \cite{Bernreuther91,Barr93}).
Measurements of the electron EDM are worthwhile as in the
future they may be able to distinguish between these models.
Direct
measurements of the electron EDM are difficult due to the
electron's charge (see, e.g., \cite{Commins94}), so 
results for the electron EDM are obtained from atomic 
EDM measurements instead.
The present limit on the electron EDM is
$|d_e|< 4 \times 10^{-27} e \mbox{ cm}$, from an
experiment on the
EDM of the Tl atom \cite{Commins94}.
A summary of the results of atomic EDM measurements can
be found in Table 6.2 of Ref.\ \cite{CPVWS}.

An atomic EDM can be induced
by the presence of an electron EDM (as well as by other
mechanisms, such as $T$- and $P$-violating electron-nucleon
or nucleon-nucleon interactions, and nucleon EDMs). 
This
allows us to obtain experimental results for the
electron EDM by measuring the atomic EDM.
The atomic EDM ($d_A$) induced by the electron EDM ($d_e$)
would be proportional to $d_e$, and the ratio
$R = d_A / d_e$ for a particular atom
is known as the enhancement factor. As was first noted
by Sandars \cite{Sandars65}, the atomic EDM of a
heavy atom can be many times larger than the electron
EDM; in fact, $R$ is of the order of $Z^3 \alpha^2$ times
a relativistic factor ($\sim 3$ in heavy atoms), where
$Z$ is the atomic number and $\alpha = 1/137$ (see, e.g.,
\cite{Flambaum76,PNCIAP,CPVWS}). To convert experimental results
for
atomic EDMs to results for electron EDMs the value of
the enhancement factor $R$ is needed.
Summaries of results for various enhancement factors
can
be found in Ref.\ \cite{Commins94} and Table 6.1 of Ref.\
\cite{CPVWS}.
In this work
we do accurate calculations of $R$ for the Fr and Au atoms.

The $T$- and $P$-odd interaction between the
EDM of an electron and the electric field of the nucleus
($-d_e \gamma_0 {\bf \Sigma} \cdot {\bf E}$)
results in an admixture of the
ground state of the electron with excited
states of opposite parity, according to perturbation theory.
On one side of the atom the ground state and
excited state wave functions 
will have the same sign, while on the other side they will
have opposite signs. Therefore the total wave function
will be larger on one side of the atom, hence the electron
will be more likely to be there and so the atom
will have an EDM. An expression for this atomic EDM, and hence
$R$, can be presented in the following form
(see, e.g., Refs.\cite{Flambaum76,PNCIAP,CPVWS}):
\begin{equation}
R = \frac{d_A}{d_e}
= 2 e \sum_n \mbox{Re} \frac{\langle 0 | z | n \rangle
\langle n | (\gamma_0 - 1) {\bf \Sigma} \cdot
{\bf E} | 0 \rangle}{E_0 - E_n}
\label{eenhfacexp}
\end{equation}
(for an atom with one valence electron), where
$| 0 \rangle$ is the unperturbed ground state,
$\{ | n \rangle \}$ is the set of states with which it
is mixed
(including unbound, continuum states for which the sum
should be replaced by an integral),
$-e$ is the charge on the electron,
${\bf E}$ is the electric field produced by the nucleus
[$= Z e {\bf r} / r^3$; the main contribution
to the second matrix element in (\ref{eenhfacexp}) comes
from short distances, where the electric field of the
nucleus is unscreened],
$z = {\bf r}_z$, and $\gamma_0$ and
${\bf \Sigma}$ are the normal matrices of relativistic
quantum mechanics.
The ground state of the valence electron in Fr is $7s$,
while in Au it is $6s$.
The operator $(\gamma_0 - 1) {\bf \Sigma} \cdot {\bf E}$ is a
pseudoscalar and so it can only mix states having opposite
parity and the same total angular momentum. Therefore the ground
state is only mixed with $p_{1/2}$ states.

Equation (\ref{eenhfacexp}) can be rewritten as
\begin{equation}
R = -\frac{4 Z \alpha}{3} \mbox{Re} \sum_n
\frac{\int_0^{\infty} [f_s^*(r) r f_{np}(r)
+ g_s^*(r) r g_{np}(r)] r^2 Q(r) \, dr
\int_0^{\infty}
g_{np}^*(r) r^{-2} g_s(r) r^2 P(r) \, dr}{E_0 - E_{np}}
\label{eefint}
\end{equation}
by using the following expression for the electron's
relativistic
wave function (see, e.g., \cite{RelQuantTheory}):
$\psi_{nljm} = (f_{njl}(r) \Omega_{jlm}, -i g_{njl}(r)
(\bbox{\sigma} \cdot {\bf r} / r) \Omega_{jlm})^T$,
where $f$ and $g$ are radial wave functions ($f_s$ refers to the
ground state and $f_{np}$ the $p_{1/2}$ excited states) and
$\Omega_{jlm}$ is a spherical spinor (an eigenfunction
of $\hat{j}^2$ and $\hat{j}_z$).
Computer generated
wave functions were used to calculate the integrals in 
Eq.\ (\ref{eefint}) (where available, previously
determined values were used).
These wave functions were obtained using the relativistic
Hartree-Fock method.
The factors $P(r)$ and $Q(r)$ take into account
the screening of the nuclear electric field by electrons
and core polarization corrections (for the non-direct
core polarization contribution these factors actually
become non-local operators).
We also took into account correlation corrections to the
wave functions.
The many-body
perturbation theory methods that we used are
described in 
Refs.\ \cite{Dzuba87,Dzuba89a,Dzuba89b,Dzuba89c,Dzuba95}.
As a test, we also performed calculations using a semi-empirical
method \cite{Flambaum76} that does not require computer
calculations. The results were in good agreement with the
numerical calculations.

For Fr, we used the experimental value of the
$7s$-$7p_{1/2}$ radial integral of $r$
[the first integral in (\ref{eefint})] that was determined
in \cite{Simsarian98}: $-5.238(10)$, in units
of the Bohr radius (this compares well with the
calculated value in \cite{Dzuba95}: $-5.241$).
For the
$7s$-$8p_{1/2}$ radial integral of $r$ we used the
value calculated in \cite{Dzuba95}
(we used the most complete many-body calculation value
denoted by ``Brueckner plus non-Brueckner''
in Table IV of this work), with
an estimated accuracy of $3\%$.
The $7s$-$9p_{1/2}$ and $7s$-$10p_{1/2}$ radial integrals of $r$
were calculated by us,
as were the $7s$-$np_{1/2}$ values of the second integral
in (\ref{eefint}) (all with an estimated accuracy of 3\%). We used
the values of the $7s$, $7p_{1/2}$, $8p_{1/2}$, and
$9p_{1/2}$ energy levels listed in \cite{D95andrefth},
while we calculated the $10p_{1/2}$ energy level
ourselves. We truncated the summation in (\ref{eefint})
for the discrete states at the $10p_{1/2}$ state, as the
remainder of this series is very small (it gives a contribution
to the enhancement factor $\sim 3$, i.e., $0.3\%$).
For the unbound, continuum states all integrals were calculated
in the present work. For these we did not take
into account screening, core
polarization or correlation corrections, and so the
errors for these integrals are larger (we
estimate a $50\%$ error for the whole continuum
contribution),
though this does not have
an excessively large effect on the final error
as the contribution of the continuum states to
the enhancement factor
is small ($=30$). The final value of the enhancement factor
for Fr is 910, with an estimated 5\% error.

For Au, we used the energy levels
listed in \cite{Moore}.
Using the experimental result for the
oscillator strength for the $6s$-$6p_{1/2}$ transition
in \cite{Hannaford81}, we obtained the value of
the $6s$-$6p_{1/2}$ radial integral of $r$: $-2.16(2)$.
All of the other radial integrals were
calculated in the present work, with an
estimated accuracy of 10\% for
the discrete states.
Once again, we estimate the error for the continuum
contribution as $50\%$
[the continuum contribution was again small ($=20$)].
We truncated the summation over discrete states
at the $8p_{1/2}$ state, with the remainder
of the series giving a contribution $\sim 1$ ($0.3\%$)
to the enhancement
factor. The final value of the enhancement factor for Au
is 260, with an estimated 15\% error.

These results are in reasonable agreement with the previously
determined estimates of the enhancement factors:
$\approx 1150$ for Fr \cite{SandarsFr}
and $\approx 250$ for Au \cite{Johnson86}.

\end{document}